\documentclass[prd,twocolumn,preprintnumbers,nofootinbib,tightenlines,superscriptaddress]{revtex4}

\usepackage{lipsum}

\pdfoutput=1
\usepackage[english]{babel}
\usepackage{amsmath,amssymb,amsfonts,bm,bbm,slashed,subdepth}
\usepackage{graphicx}
\usepackage[sort&compress]{natbib}
\usepackage{xcolor}
\usepackage[normalem]{ulem}
\usepackage{hyperref}
\usepackage{cleveref}
\definecolor{red}{rgb}{1.0, 0, 0}
\usepackage{enumerate}
\usepackage{epsfig, subfigure}
\usepackage{setspace}
\usepackage{booktabs, tabularx}
\usepackage{units}
\usepackage{placeins}
\usepackage{multirow}
\usepackage{mathtools}

\newcommand{\Lag}{\mathcal{L}}

\newcommand{\prn}[1]{ \left(  #1 \right) }

\newcommand{\al}[1]{\begin{align} #1 \end{align}}

\definecolor{myred}{rgb}{0.6078431372549019,0.11372549019607843,0.12549019607843137}
\definecolor{myblue1}{rgb}{0.16791, 0., 0.301671}
\definecolor{myblue2}{rgb}{0.28235299999999997, 0.1497248, 0.6790623333333333}
\definecolor{myblue3}{rgb}{0.26669366666666666, 0.550462, 0.926485}
\definecolor{myblue4}{rgb}{0.6711543333333333, 0.814616, 0.9359733333333333}
\definecolor{mypurp}{rgb}{0.761959, 0.470832, 0.940597}

\begin{document}

\preprint{FERMILAB-PUB-20-365-AE-T}
\title{Resonant Self-Interacting Dark Matter from Dark QCD}

\author{Yu-Dai Tsai}
\email{ytsai@fnal.gov}
\affiliation{Fermilab, Fermi National Accelerator Laboratory, Batavia, IL 60510, USA}
\affiliation{University of Chicago, Kavli Institute for Cosmological Physics, Chicago, IL 60637, USA}

\author{Robert McGehee}
\email{robertmcgehee@berkeley.edu}
\affiliation{Department of Physics, University of California, Berkeley, CA 94720, USA}
\affiliation{Ernest Orlando Lawrence Berkeley National Laboratory, Berkeley, CA 94720, USA}

\author{Hitoshi Murayama}
\email{hitoshi@berkeley.edu} \email{hitoshi.murayama@ipmu.jp, Hamamtsu Professor}
\affiliation{Department of Physics, University of California, Berkeley, CA 94720, USA}
\affiliation{Kavli Institute for the Physics and Mathematics of the
  Universe (WPI), University of Tokyo,
  Kashiwa 277-8583, Japan}
\affiliation{Ernest Orlando Lawrence Berkeley National Laboratory, Berkeley, CA 94720, USA}

\begin{abstract} 
We present models of resonant self-interacting dark matter in a dark sector with QCD, based on analogies to the meson spectra in Standard Model QCD. For dark mesons made of two light quarks, we present a simple model that realizes resonant self-interaction (analogous to the $\phi$-$K$-$K$ system) and thermal freeze-out. We also consider asymmetric dark matter composed of heavy and light dark quarks to realize a resonant self-interaction (analogous to the $\Upsilon(4S)$-$B$-$B$ system) and discuss the experimental probes of both setups. Finally, we comment on the possible resonant self-interactions already built into SIMP and ELDER mechanisms while making use of lattice results to determine feasibility.
\end{abstract}
\maketitle
\section{Introduction}
The study of dark matter (DM) has been one of the most important topics in particle physics, astrophysics, and cosmology.
Although there is overwhelming evidence of DM, we know next to nothing about its nature. Observations involving halo or subhalo structures \cite{Moore:1994yx} may shed light on this mystery.
Historically, core versus cusp \cite{Dubinski:1991bm,Navarro:1995iw,Flores_1994,Moore:1994yx}, too-big-to-fail \cite{Boylan_Kolchin_2011}, and diversity problems \cite{Oman_2015} have indicated the potential existence of DM self-interaction (see, e.g., \cite{Tulin_2018}), although baryonic feedback \cite{Navarro_1996,Gelato_1999,Binney_2001,Gnedin_2002} provides an alternative explanation of these small-scale puzzles.

The Bullet cluster~\cite{Clowe:2003tk,Markevitch:2003at,Randall:2007ph}, along with halo shape observations~\cite{Rocha:2012jg,Peter:2012jh}, set an upper bound on DM self-interactions around $\sim\rm cm^2/g$. Given that a larger cross-section could be preferable for smaller-scale halos \cite{Kaplinghat:2015aga}, introducing a velocity dependent self-interaction to explain the small-scale structure issues is well motivated. 

The preferred DM self-interaction strength is near that of nuclear interactions \cite{Tulin_2018}. Thus, it is interesting to consider a QCD-like theory in which such strength of interaction emerges. Additionally, the simplest way to achieve such velocity dependence solely in the dark sector is via resonant scattering \cite{Chu:2018fzy}. Suppose there is a resonance in the DM self-interactions just above the threshold of twice its mass. Then this resonant self-interacting DM (RSIDM) may miss this resonance in systems with large velocity dispersions, such as clusters of galaxies, while it may frequently hit the resonance in systems with small velocity dispersions, such as dwarf galaxies. This would lead to cross-section enhancement at small velocities, yielding the desired velocity dependence. This solution typically requires the resonance have a mass $\prn{10^{-6}-10^{-4}} m_{\text{DM}}$ above twice the DM mass.

In this paper, we will consider multiple models with mediators just above the threshold which explains such resonances. To achieve these resonances, we need to look no further than Standard Model (SM) QCD in which many cases of such resonances exist naturally. Perhaps the most famous example of near-threshold resonance is in the triple-$\alpha$ reaction in stellar burning, $\alpha\alpha \rightarrow {}^8{\rm Be}$, $\alpha{}^8{\rm Be} \rightarrow {}^{12}{\rm C}^*$ (7.66~MeV $0^+$ excited state of ${}^{12}{\rm C}$),
\begin{align}
    \frac{m({}^8{\rm Be})-2m(\alpha)}{m({}^8{\rm Be})} &= 0.000012, \\
    \frac{m({}^{12}{\rm C}^*)-m({}^8{\rm Be})-m(\alpha)}{m({}^{12}{\rm C}^*)} &= 0.000026.
\end{align}
This example is often invoked as evidence for the anthropic principle~\cite{Barrow:1988yia,CCref}. Even though they are less pronounced, there are numerous examples of near-threshold resonances in QCD, such as
\begin{align}
    \frac{m(\phi) - 2m(K^0)}{m(\phi)} &= 0.024, \\
    \frac{m(D^{0*}) - m(D^0)-m(\pi^0)}{m(D^{0*})} &= 0.0035, \\
    \frac{m(B_{s1}) - m(B^*) - m(K^0)}{m(B_{s1})} &= 0.0011, \\
    \frac{m(\Upsilon(4S)) - 2m(B^0)}{m(\Upsilon(4S))} &=0.0019.
\end{align}

\begin{figure}[t!]
    \centering
    \includegraphics[
    width=\columnwidth]{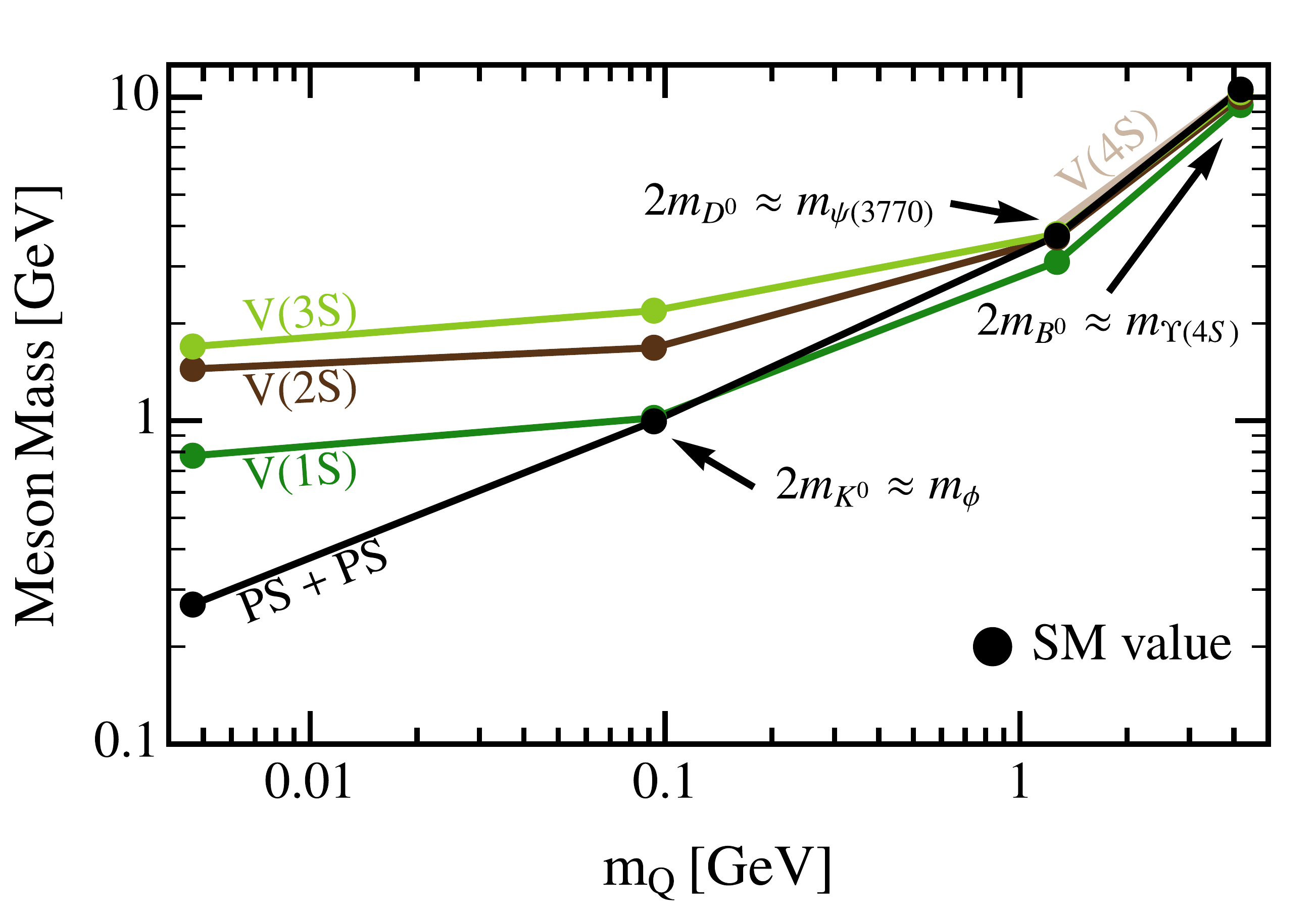}
    \caption{
    A selection of the SM meson spectrum as a function of the larger quark mass in each meson, $m_Q$. Extrapolations of twice the pseudoscalar meson mass (PS+PS), of the first vector meson mass (V(1S)), of the second vector meson mass (V(2S)), of the third vector meson mass (V(3S)), and of the fourth vector meson mass (V(4S)) are shown. For $m_Q=m_d$, we show $\pi^0$ as well as the average masses of the first three $\rho$ and $\omega$ states. For $m_Q=m_s$, we show $K^0$ and the first three $\phi$'s. For $m_Q=\{m_c,m_b\}$, we show $D^0$ and $B^0$ as well as the first four $\psi$ and $\Upsilon$ states, respectively.
    }\label{fig:resonance}
\end{figure}

Some of these illustrative near-resonances are shown in Fig.\ref{fig:resonance}.  Most examples are actually not pure accidents: 
QCD dynamics require there to be such near-threshold resonances. In a heavy-light meson $(Q\bar{q})$, its mass is essentially the sum of the heavy quark mass $m_Q$ and the effect of the strong interaction $\sim \Lambda_{QCD}$. On the other hand, for the heavy-heavy meson $(Q\bar{Q})$, its mass is twice the heavy quark mass $2m_Q$ and the effect of binding. In the limit $m_Q \gg \Lambda_{QCD}$, it is clear $m_{Q\bar{Q}} \approx 2 m_{Q\bar{q}}$ is the zeroth-order approximation. To be more precise, we need to understand the quarkonium potential, which is discussed in Section III. On the other hand, the mass splitting between $D^*$ and $D$ is due to the hyperfine interaction between magnetic moments and is approximately $\sim \Lambda_{QCD}^2/m_Q$ which is totally unrelated to $m_\pi \approx (m_q \Lambda_{QCD})^{1/2}$. We consider this example to be a pure accident.

In the following sections, 
we discuss three specific scenarios. First, we outline a model with 2 light quarks, with one much heavier than the other, in which dark ``kaons'' freeze-out to the correct relic abundance and the resonance is analogous to $K^{+} K^{-} \rightarrow \phi$. We then discuss an asymmetric DM model in which DM particles are mesons with one heavy and one light quark and the resonance is similar to $B^{0} \overline{B}^{0} \rightarrow \Upsilon(4S)$.
The closeness to threshold $\Delta \equiv 1-2m_{\text{PS}}/m_{\text{V}}$ in both must be quite significant, where $m_{\text{PS}}$ is the mass of the pseudoscalar meson and $m_{\text{V}}$ is the mass of the vector meson.
Finally, we describe a model directly based on the Strongly Interacting Massive Particle (SIMP) framework discussed in \cite{Hochberg:2014kqa} and use lattice results to determine the parameters for the resonance.

With our discussions of QCD mesons and resonances complete, future references to quarks (\emph{e.g.} $u$) and mesons (\emph{e.g.} $K$) in this paper will refer to dark sector analogues to the SM states unless otherwise noted.

\section{Light Quark Model}
\label{sec:light-quark}

For the model outlined in this section, we assume a QCD-like gauge theory $SU(3)_D$ in the dark sector. DM is comprised of dark ``kaons''\footnote{Even though we call the DM mesons kaons, they are the lightest $SU(3)_D$ states. This name was chosen since $m_{s} \gg m_{u}$ and we are motivated by the near-threshold resonance in the SM process $K^{+} K^{-} \rightarrow \phi$.} composed of two dark quarks with masses much smaller than the dark QCD scale, labeled $u$ and $s$, with $m_{s} \gg m_{u}$. 
The quarks are charged under a dark $U(1)_D$ as \(u(+1)\) and \(s(0)\) which is broken, resulting in a massive dark photon $A_D$. We also assume a kinetic mixing between $U(1)_D$ and $U(1)_{\text{EM}}$ of the form $\Lag \supset 1/2 \cdot \epsilon F_{\mu \nu}F_D^{\mu \nu}$. 

{\bf DM self-interactions - }The desired resonant self interaction is provided by the dark \(\phi\) exchange saturating the Breit-Wigner cross section in the \(P\)-wave. We assume \(\Delta \sim 10^{-7.8}\) for these dark mesons \cite{Chu:2018fzy}. We also need \(\frac{\sigma _0}{m_{\rm DM}}\sim 0.1 \frac{\text{cm}^2}{g}\) in order for the low-velocity limit of the self-interaction cross-section to fit small-scale structure observations \cite{Chu:2018fzy} (also see discussions around our Eq. \ref{eq:sigma_0}). 
Thus, we calculate the 4-kaon interaction in the dark sector.

We define 
\(U=e^{2i \Pi \left/f_{K }\right.}\), \(\Pi =K ^aT^a=\frac{1}{2}K ^a\tau ^a\), \(2 \text{Tr}(\Pi^2 )=K ^aK ^a\), $f_K$ is the dark kaon decay constant. 
First, consider the non-derivative couplings. The relevant Chiral Lagrangian terms are:
\begin{align}
\mathcal{L}
=&\frac{1}{2}\frac{m_{K}^2f_{K }^2}{m_u+m_s}\text{Tr}\left[U^{\dagger}\left(
\begin{array}{cc}
 m_u & 0 \\
 0 & m_s \\
\end{array}
\right)+\left(
\begin{array}{cc}
 m_u & 0 \\
 0 & m_s \\
\end{array}
\right)U\right]
\\
&\supset-m_{K }^2K^+K^- +\frac{m_{K }^2}{6f_{K }^2} \prn{K^+K^-}^2
\end{align}
The relevant derivative couplings are
\begin{align}
\mathcal{L}=&
\frac{f_{K }{}^2}{4}\text{Tr} \partial _{\mu }U^{\dagger }\partial ^{\mu }U
\\
= & \partial _{\mu }K ^+\partial ^{\mu }K ^-
-\frac{2m_{K }{}^2}{3f_{K }{}^2}(K^+ K^-)^2 \nonumber
\\
&-\frac{1}{2f_{K }{}^2}\left(K ^+K ^-\right)\partial _{\mu
}\partial ^{\mu }\left(K ^+K ^-\right)  + O\left(K ^6\right) \nonumber
\end{align}

We assume \(K ^0\) is heavier than \(K ^{\pm }\) by $\sim $10$\%$ (which can be induced by the $L_7$ term in the Chiral Lagrangian \cite{Pich:1995bw,Scherer:2002tk,Kubis:2007iy}), so that only the \(K ^{\pm }\) states make up DM. From here on, we define $m_K = m_{K^{\pm}}$ to be the masses of the dark charged kaons.
The neutral kaon is unstable and cannot be a DM candidate because it can decay into, for example, 4 electrons, through an off-shell dark photon. In halos today, there are only \(K ^{\pm }\) interactions. After taking into account the derivative terms, the self-interaction cross-section for $K ^+K ^-\to K ^+K ^-$ is
\begin{align}
\sigma_{K^+ K^-}=\frac{1}{16\pi }\frac{m_{K}^2}{f_{K }^4}.
\end{align}
To match the fitted low-velocity limit of the self-interaction cross-section \cite{Chu:2018fzy}, we set
\begin{align}\label{eq:sigma_0}
\frac{\sigma_{0}}{m_{\rm DM}} = \frac{1}{2} \frac{\sigma_{K^+ K^-}}{m_K}\simeq 0.11^{+0.10}_{-0.05}\;\rm cm^2/g.
\end{align}
$m_{\rm DM}=m_K$ is the DM mass and $\sigma_0$ is the low-velocity limit of the DM self-interaction cross section.

\begin{figure}[t!]
    \centering
    \includegraphics[
    width=\columnwidth]{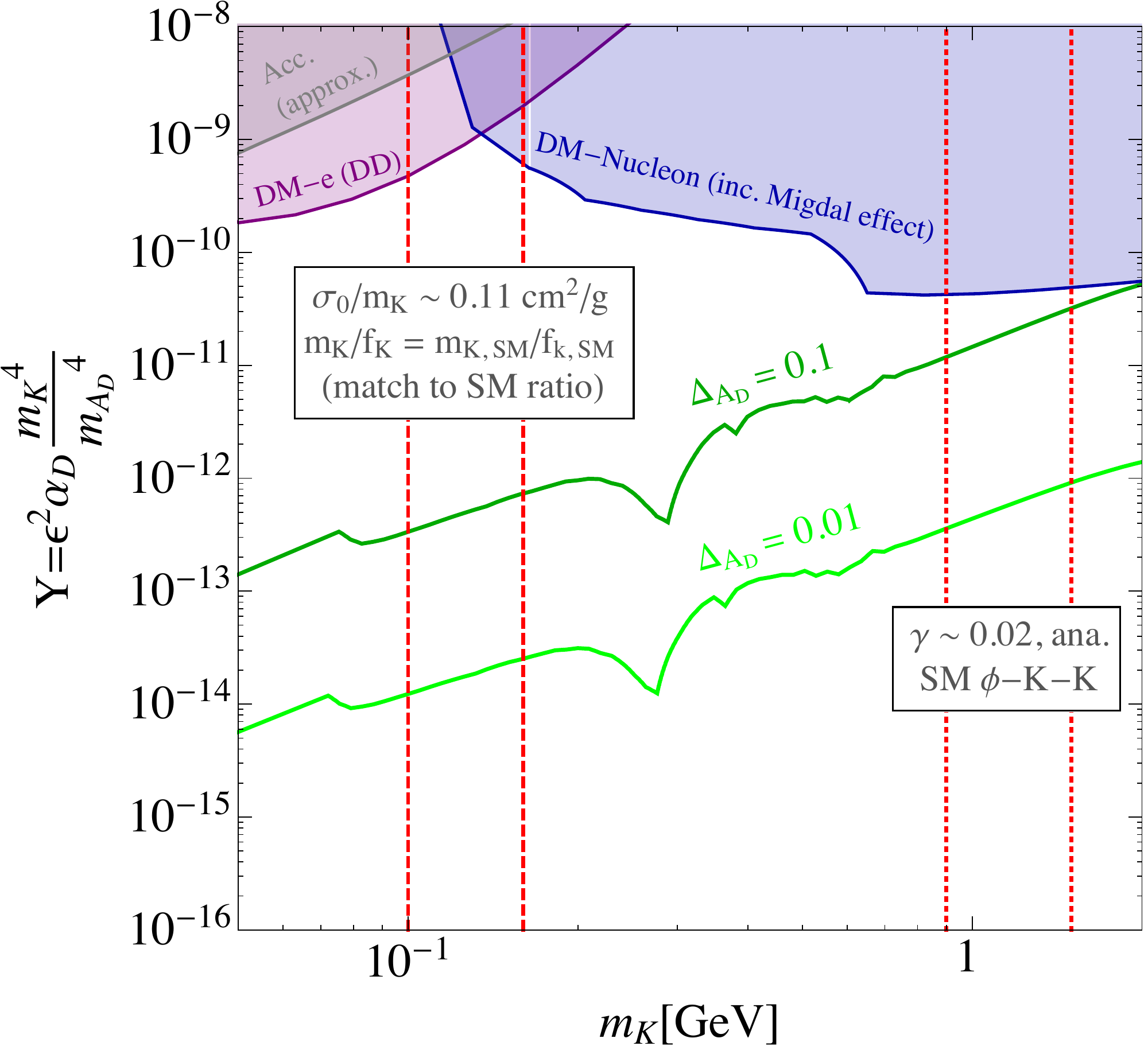}
    \caption{
The most motivated mass ranges for resonant self-interaction, analogous to the $\phi$-$ K$-$K$ system discussed in the text, are enclosed by red dashed and dotted lines.
The green curves give the correct relic abundance with $\Delta_{A_D}=({m_{A_D}^2- 4 m^2_{\rm DM})/4 m^2_{\rm DM}}$ = 0.1 and 0.01, reproduced from \cite{Feng:2017drg}. 
The purple regime is constrained by DM-electron direct detection, the gray regime is the
approximate accelerator bound (see text for discussions), and the blue region is constrained by the DM-nucleon scattering (including the Migdal effect).
}\label{fig:LDM}
\end{figure}

Fig.~\ref{fig:mvf} in the Appendix shows the parameters which give the correct $\sigma _0/m_{\rm DM}$ for RSIDM. This fixes the relation between $m_K$ and $f_K$, $f_K \sim \left(0.07\pm 0.01 \right)\rm GeV \;\left(\frac{m_K}{{\rm GeV}}\right)^{1/4}.$
If we match this to the SM ratio of $m_K/f_k \sim 0.32$ \cite{Tanabashi:2018oca}, 
we get $m_K\sim 100 - 160\; \rm MeV$ for the dark kaon (this will be referred to as as region I).
On the other hand, if we consider the SM $\phi$-$K$-$K$ system, its $\gamma \sim 0.02$ and $m_K \sim 0.9 - 1.5$ GeV (as listed in Table \ref{table:1}, this will be referred to as as region II). We delineate the ranges of $m_K$ which correspond to each of these two assumptions in Fig.~\ref{fig:LDM}. Even though these regions do not overlap, one could consider a different gauge group or simply a different $N_c$ (see appendix for more discussions). For example, the regions could move closer\footnote{Even though the large-$N_c$ limit cannot be trusted, we sketch how these regions I and II come closer for $N_c =2$ in the Appendix.} for $N_c =2$. 

The DM self-interaction mediated by the dark photon $A_D$ is suppressed as $\left(m_{K}/m_{A_D}\right)^4$, and the interaction strength is much smaller than that of 4-meson interaction, so can be neglected in this consideration.

{\bf Freeze-out - }Here we consider the process that sets the DM relic abundance. 
We assume $A_D$ is heavier than \(K ^{\pm }\). 
Since $A_D$ is heavier than \(K ^{\pm}\), before \(K^0\) decays (suppressed by one-loop, \(\epsilon^4\), and \(m_{A_D}^{-8}\)), it annihilates via \(K ^0K ^0\to K ^+K
^-\). The annihilation \(K ^+K ^-\to A_D A_D\to e^+e^-e^+e^-\) can also happen, but it is suppressed by \(\epsilon ^4\), much smaller than the freeze-out cross section.
The primary freeze-out process we consider is thus \(K ^+K ^-\to A_D \to\) SM. 

The generic choice of $m_{A_D}$ and $m_K$ is mostly excluded in our parameter region of interest. However, one can invoke another resonance to open up the parameter space. In addition to the resonance in self-interactions induced by the vector meson, one can also arrange the dark photon mass so that it goes on resonance for the freeze-out process, to allow smaller $Y$ to produce the correct relic abundance and avoid accelerator as well as direct-detection constraints \cite{Feng:2017drg,Izaguirre:2015yja,Ibe:2008ye}. We define $\Delta_{A_D} \equiv \frac{m_{A_D}^2- 4 m^2_K}{4 m^2_K}$. 
In Fig. \ref{fig:LDM}, we show the $A_D$ resonant cases with $\Delta_{A_D}$ = 0.1 and 0.01, along with constraints from direct-detection \cite{Aprile:2019jmx,Baxter:2019pnz,Aprile:2019xxb,Barak:2020fql,Amaral:2020ryn} and accelerator experiments \cite{Aguilar-Arevalo:2018wea, NA64:2019imj, Lees:2017lec, Fabbrichesi:2020wbt,Berlin:2020uwy,Krnjaic:2019dzc}. We assume $m_{A_D}= 2 m_K$ for the direct-detection constraints and rescale the accelerator constraints accordingly.\footnote{However, the accelerator constraints should be modified accordingly for different values of $\Delta_{A_D}$. As discussed in \cite{Berlin:2020uwy}, both the visible searches for DM and invisible searches for $A_D$ would be modified in the resonance regime because $A_D$ would have non-negligible branching ratios to SM particles for small $\Delta_{A_D}$. This would involve a re-analysis of experimental data beyond the scope of this work.} 

{\bf CMB and halo constraints - }For \(m_{K }\sim 100\) MeV, the relevant CMB constraint on DM annihilations is approximately \(\langle \sigma v\rangle \lesssim 10^{-29}\text{cm}^3/\text{s}\) \cite{Essig:2013goa,Leane:2018kjk,Laha:2020ivk}. Normally, this is a problem because the annihilation cross section at freeze-out required to produce the relic density is larger: \(\langle \sigma v\rangle < 5 \times 10^{-26}\text{cm}^3\sec
^{-1}\). 
But in our case, processes like \(K ^+K ^-\to A_D \to e^+e^-\) are \(P\)-wave and is suppressed by the ratio of DM veloctities at recombination and freeze-out, \(\left(v_{\text{rec}}/v_{\text{fo}}\right)^2\approx
10^{-5}\).

On the other hand, we do need to worry about the \(P\)-wave process \(K ^+K ^-\to \phi \to e ^+e ^-\) (and other similar channels with different SM final states) because it can go on resonance during recombination and in halos today. 
To be relevant for small-scale structure problems, we need \(K ^+K ^-\to \phi \to K^+ K^-\) to have 
$\langle \sigma  v\rangle /m \sim 10^2 \;\text{cm}^2/ \text{g}\times
\text{km} / \text{s},$ and thus
$ \langle \sigma  v\rangle \sim 1.7 \times 10^{-18}\left(\frac{m}{100\;\text{MeV}}\right) \text{cm}^3/\text{s}$ \cite{Chu:2018fzy}. 
\(K ^+K ^-\to \phi \to e ^+e ^-\) and \(K ^+K ^-\to A_D\to e^+e^-\) are related by the branching fraction \(\phi \to A_D \to e^+e^-\). In QCD, \(\phi \to e^+e^-\) branching fraction is a small value, \(3\times 10^{-4}\), due to the OZI rule. 
Since \(s\) has no charge, it couples to $A_D$ only through a departure from the ideal mixing. In SM QCD, this is only at the 1$\%$ level, leading to a suppression by \(10^{-4}\). On top of both, it is suppressed by \(\epsilon ^2\) as well as $\left(m_{\phi}^2/m_{A_D}^2\right)^2$. The total suppression is roughly
\begin{align}
&\text{BR}\left(\phi \to A_D \to e^+e^-\right)\\
& \approx 3\times 10^{-4}\frac{\alpha _D}{\left(\frac{1}{3}\right)^2\alpha }(1\%)^2\left(m_{\phi
}/m_{A_D}\right)^4\epsilon ^2
\\
& \approx 3\times 10^{-16}\frac{\alpha _D}{\left(\frac{1}{3}\right)^2\alpha }\left(m_{\phi
}/m_{A_D}\right)^4\left(\frac{\epsilon}{10^{-4}}\right)^2.
\end{align}
Thus, this process is safe from CMB and galactic halo constraints, even if $s$ is not neutral under $U(1)_D$. For simplicity, one may still take $s$ neutral and assume ideal mixing so that \(K ^+K ^-\to \phi \to e ^+e ^-\) is forbidden.
The argument is true for the other similar relevant channels with different SM final states.

\section{Heavy Quark Model}

For the model discussed in this section, we want the near-threshold resonance to emerge directly from the theory.
We consider one light quark $u$ and two heavy quarks $c$ and $b$ and assume the $c$ and $b$ abundances are fixed by their asymmetries, $n_{c} = n_{\bar{b}}$. There are many ways to populate asymmetric DM (see, \emph{e.g.} ~\cite{Davoudiasl:2012uw,Petraki:2013wwa,Zurek:2013wia} and references therein) which will work for this GeV scale DM\footnote{It is possible to generate both the DM and baryon asymmetries in models with a dark QCD, \emph{e.g.} \cite{Hall:2019rld}.}. So, we remain agnostic about the origin of the asymmetry. We also assume the heavy quarks have a common mass, $m_Q$, and refer to either heavy quark as $Q$.  This assumption is not necessary for the successful phenomenology, and is made only for the simplicity of discussions.  The resonance is $D^{0} (c\bar{u}) B^{+} (u\bar{b}) \rightarrow \Upsilon(c \bar{b})(nS)$ for some excited level $n$ and $m_D=m_B$ is the DM mass for these heavy-quark mesons. Despite being motivated by the presence of the heavy quarks, this resonance requires some level of accident which we proceed to estimate. 
The relic abundance of the DM particles, $D^{0}$ and $B^{+}$, are set by the asymmetry of $n_{c}$ an  $n_{\bar{b}}$.

We introduce a massive dark photon $\gamma'$ corresponding to a broken $U(1)'$ dark gauge group which the lightest pseudoscalar dark meson, $\pi (\bar{u} u)$, decays through.\footnote{Massive dark photons can efficiently transfer entropy between the Standard Model and dark sectors before neutrino decoupling and are especially useful when there are many degrees of freedom from a dark QCD (\emph{e.g.} ~\cite{Harigaya:2019shz,Koren:2019iuv,Hall:2019rld}).} We assume a similar coupling as the SM $\pi^0$ to two photons and that the decay proceeds through a heavy-fermion loop. 
Note that $\gamma'$ here is different from the dark photon $A_D$ introduced in Sec. II since $\gamma'$ decays entirely to visible SM particles. We assume a kinetic mixing between $U(1)'$ and $U(1)_{\text{EM}}$ of the form $\Lag \supset 1/2 \cdot \epsilon F^{\mu \nu} F'_{\mu \nu}$.

{\bf Heavy-light meson and quarkonium spectrum- }Following the discussion of \cite{Quigg:1977xd,Quigg:1977dd,Quigg:1997bx}, interactions of heavy quarks can be described by the non-relativistic Schr\"{o}dinger equation. 
The $c \bar{b}$ bound states 
have the logarithmic potential 
 \begin{align}
\begin{split}
V(r) = C \ln (r/r_0) \, ,
\end{split}
\end{align}
where $C$ is a parameter that can be calculated in lattice QCD and $r_0$ is the distance at which the log potential is equal to the threshold necessary for $\Upsilon(c \bar{b})$ to decay into $D^{0} (c\bar{u}) + B^{+} (u\bar{b})$.
The level spacing of these quarkonium excited states are independent of $m_Q$ (see Eq. 8 of \cite{Quigg:1977xd}):
 \begin{align}
\begin{split}
m_{\Upsilon(nS)}-m_{\Upsilon(1S)} \approx C \ln \left(\frac{4n}{3}\right)
\label{eq:level_nS}
\end{split}
\end{align}
in the large $n$ limit. The mass splitting is 
 \begin{align}
\begin{split}
\Delta_n &\equiv m_{\Upsilon(nS)}-m_{\Upsilon((n-1)S)}\\
& =  C \left[\frac{1}{n} +  \mathcal{O}\left(\frac{1}{n^2}\right)\right].
\end{split}
\end{align}

\begin{figure}[t!]
    \centering
    \includegraphics[
    width=\columnwidth]{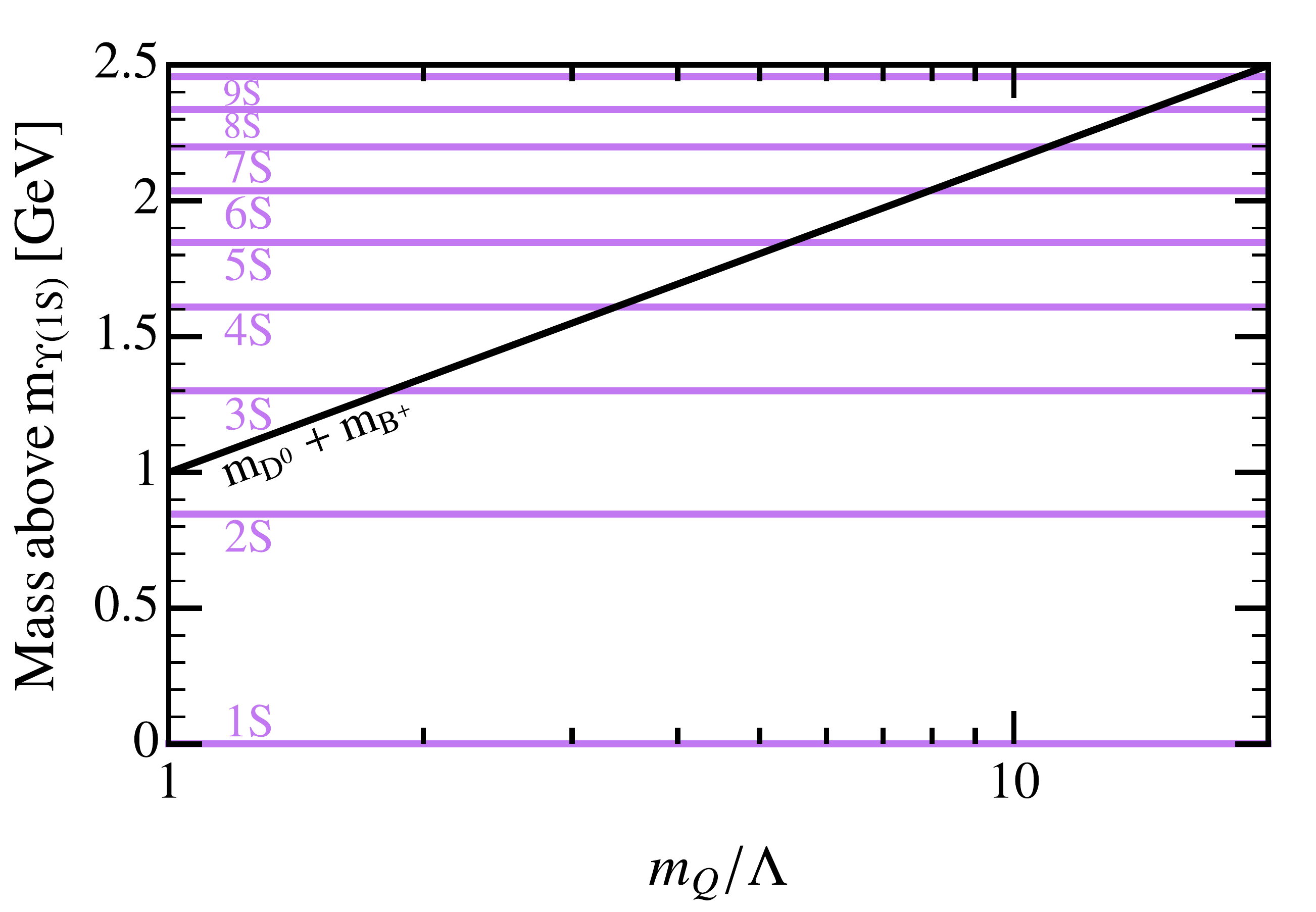}
    \caption{The crossings of the sum of heavy quark pseudoscalar meson masses and heavy quarkonium excited states for different heavy quark masses, $m_Q$.
}\label{fig:hq}
\end{figure}

The summed mass of the mesons with one heavy quark is (see, also, Eq. 6 of \cite{Quigg:1977xd})
 \begin{align}
\begin{split}
m_D+m_B -m_{\Upsilon(1S)} = A + \frac{1}{2} C \ln \left(\frac{m_Q}{\Lambda}\right), 
\label{eq:level_DB}
\end{split}
\end{align}
assuming $m_Q \gg \Lambda$, where $\Lambda$ is the dark confinement scale.\footnote{If $m_c \ne m_b$, the expression in Eq.~\ref{eq:level_DB} would be more complicated, but the behavior in Fig.~\ref{fig:hq} would remain.} 
The intersection of the summed scalar meson masses (black) with the different heavy quarkonium excited states (\textcolor{mypurp}{purple}) is where resonance occurs as shown in Fig.~\ref{fig:hq}. 

\begin{figure}[t!]
    \centering
    \includegraphics[
    width=\columnwidth]{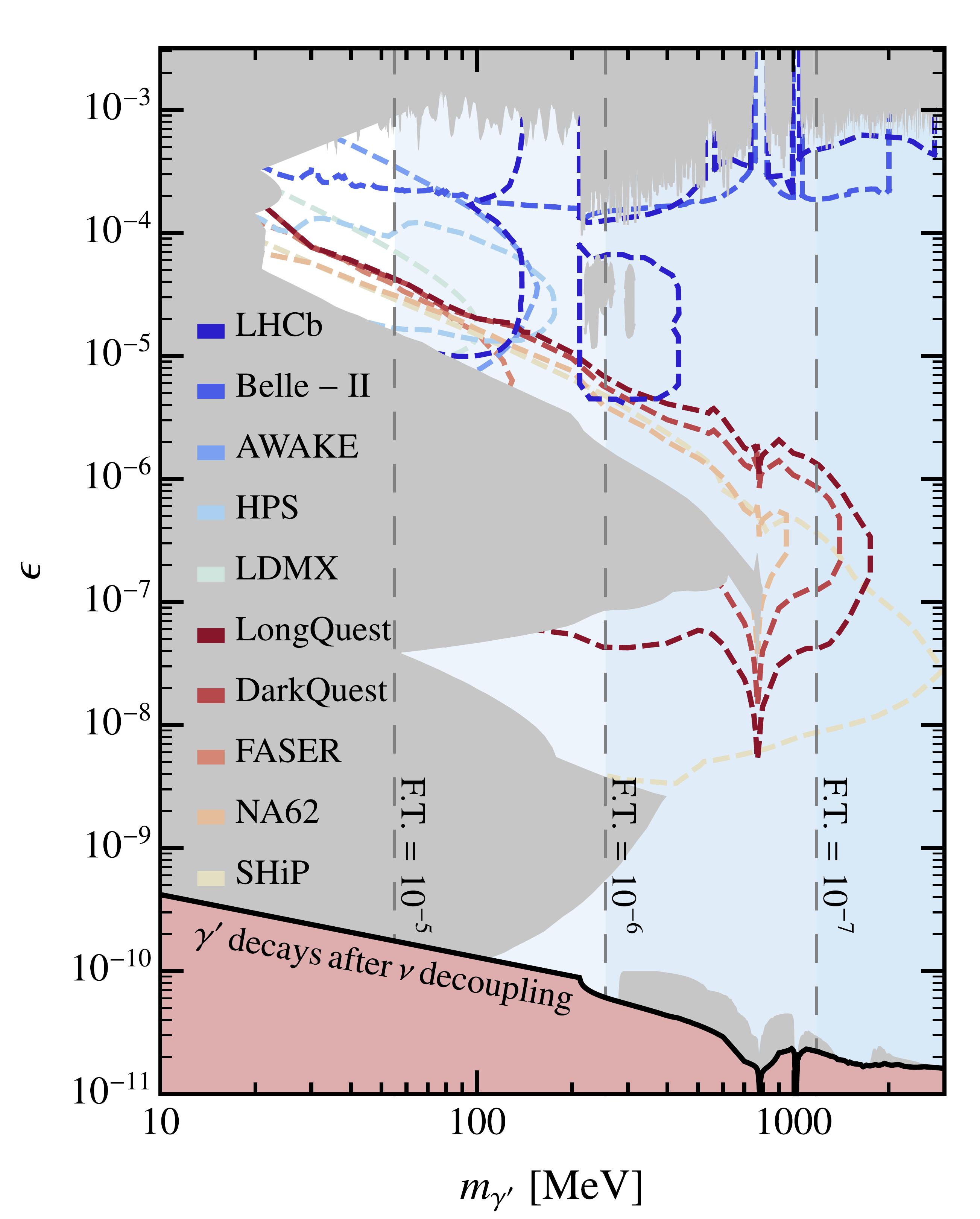}
    \caption{The parameter space in the heavy quark model in which the dark photon decays quickly enough to transfer the dark $\pi$'s entropy before neutrino decoupling. Existing constraints~\cite{Fradette:2014sza,Alexander:2016aln,Chang:2016ntp,Hardy:2016kme,Pospelov:2017kep,Banerjee:2018vgk,Aaij:2017rft,Aaij:2019bvg,Parker:2018vye,Tsai:2019mtm} are dark gray while projected sensitivities of future experiments are shown as dashed lines~\cite{Celentano:2014wya,TheBelle:2015mwa,Ilten:2015hya,Alekhin:2015byh,Ilten:2016tkc,Alexander:2016aln,Caldwell:2018atq,Berlin:2018pwi,Berlin:2018bsc,Ariga:2018uku,NA62:2312430,Tsai:2019mtm}. The dashed vertical lines show different levels of accident (F.T.) required to achieve the necessary resonant self interaction, as defined in Eq.~\eqref{eq:FT}. See the text for more details.
    }\label{fig:dp}
\end{figure}

The tuning to be on resonance can be reduced to \mbox{$\Delta \times \frac{m_Q}{\Delta_n}$}, where $\Delta$ is at the level of $10^{-7.8} $~\cite{Chu:2018fzy}. 
In the large $n$ limit, assuming the dimensionful parameters $A \sim C \sim \Lambda$ for simplicity, the $m_Q$ which allows the sum of the pseudoscalar mesons' masses to fall between the $n-1$ and $n$ levels is 
\begin{align}
\begin{split}
\label{eq:mQ}
m_Q \approx n^2 \left(\frac{4}{3 e}\right)^2 \Lambda.
\end{split}
\end{align}
The requisite level of accident (F.T.) to achieve the desired resonant self-interaction, 
\begin{align}
\begin{split}
\label{eq:FT}
\text{F.T.}\equiv\Delta \times \frac{m_Q}{\Delta_n} 
&\approx \Delta\times \left(\frac{4}{3 e}\right)^2  n^3,
\end{split}
\end{align}
can then be reduced (getting closer to an order one number).
When $n>10$, the level of accident is reduced by as much as $10^5$.

{\bf Log potential region  - }When $m_Q$ is significantly larger than $\Lambda$, 
the quark potential is Coulombic for small $n$. 
The quark potential only becomes logarithmic, as assumed above, for large enough $n$, 
which can be estimated as follows. The Bohr radius of the system is $a = 1/(\alpha_s m_Q)$, where $\alpha_s$ is the dark gauge fine structure constant. The energy levels are roughly $E_n\sim \frac{\alpha_s m_Q}{n^2}$ in the Coulombic region, so $\frac{\alpha_s m_Q}{n^2}>\Lambda$ corresponds to the log potential region.
Thus, for $m_Q \gtrsim 10\;\Lambda$ (assuming $\alpha_s\sim 1$), $n$ needs to be larger than at least 4 for the quark-system to have a logarithmic potential. This is consistent with our analysis above.

{\bf Experimental signature - }We assume the dark $\pi$ and the dark photon have the same couplings as their SM counterparts so that the former decays to the latter quickly after confinement. The dark photon must further decay to the SM to successfully transfer the excess, symmetric entropy from the dark sector prior to SM neutrino decoupling. 

As mentioned previously, we assume $n_c=n_{\bar{b}}$ for simplicity. Now, let us further assume that $n_c+n_{\bar{b}}=n_{B,\rm SM}$, where the latter is the asymmetric SM baryon number density. This could easily occur in a full model which includes a mechanism for all three asymmetries to be generated simultaneously. Requiring the asymmetric heavy-light mesons to reproduce the observed DM relic abundance yields
\begin{equation}
    m_{\rm DM}=m_p \frac{\Omega_{\text{DM}} h^2}{\Omega_{B,\rm SM} h^2},
\end{equation}
where $m_p$ is the proton mass. The DM mass $m_{\rm DM}=m_D=m_B\sim m_Q$ in the heavy-quark limit.
With Eqs.~\eqref{eq:mQ} and~\eqref{eq:FT}, we can write the required dark confinement scale as:
\begin{align}
\label{eq:LambdafrmFT}
    \Lambda \approx m_Q \prn{\frac{3 e \Delta}{4 \, \text{F.T.}}}^{2/3} 
    \sim m_p \frac{\Omega_{\text{DM}}}{\Omega_{B,\rm SM}} \prn{\frac{3 e \Delta}{4 \, \text{F.T.}}}^{2/3}.
\end{align}

To enable the dark $\pi$ to decay to a pair of dark photons, we require $2m_{\gamma '} \le m_\pi \approx \Lambda$. Thus, the upperbound on the dark photon mass is set by the level of accident we permit in Eq.~\eqref{eq:LambdafrmFT}. In Fig.~\ref{fig:dp}, we show the dark photon parameter space in which the dark pions decay to dark photons which in turn decay to SM particles fast enough. We also show the relevant current and future experimental probes.

Based on this specific dark-photon setup, the reduction of the level of accident is at best $\sim 10^3$, not the value of $10^5$ discussed below Eq. \ref{eq:FT}. However, one can consider other similar models to achieve a better reduction. For example, the dark pion can decay to completely secluded dark-sector particles (thus allowing a smaller $\Lambda$) and a better reduction of accident can be achieved.

\section{SIMP \& ELDER DM as Resonant SIDM}

In this section, we continue the theme that a natural place to expect resonances (near twice the DM mass) is in dark sectors with confining gauge groups. Two classes of such dark sectors that have their own strong motivations are Strongly Interacting Massive Particles (SIMPs)~\cite{Hochberg:2014dra} and Elastically Decoupling Relics (ELDERs)~\cite{Kuflik:2015isi}.

The SIMP mechanism achieves thermal DM which is lighter than typical WIMPs by altering the number-changing process. In typical WIMP scenarios, the DM relic abundance is set by processes such as $\bar{\chi} \chi \to \bar{f} f $, where $f$ is a SM particle. In contrast, in SIMP scenarios, the DM relic abundance is primarily set by processes in the dark sector, usually taken to be $3 \to 2$ DM processes. This alternate thermal history allows SIMPs to evade CMB bounds which typically constrain such light DM masses~\cite{Leane:2018kjk}. Since the annihilation of 3 DM to 2 leaves the latter with excess energy, SIMP scenarios also require some mediation of this extra entropy to the SM. How one connects the dark and SM sectors to ferry this entropy without spoiling the SIMP mechanism yields rich phenomenology with many interesting signatures (see, \emph{e.g.}~\cite{Lee:2015uva,Bernal:2015xba,Hochberg:2015vrg,Lee:2015gsa,Choi:2016hid,Bernal:2017mqb,Choi:2017mkk,Hochberg:2017khi,Kuflik:2017iqs,Choi:2017zww,Berlin:2018tvf,Tsai:2018qoa,Choi:2018iit,Hochberg:2018vdo,Hochberg:2018rjs}).

The ELDER scenario is similar to the SIMP mechanism: the relevant, DM-number-changing process is assumed to be one in the dark sector such as a $3 \to 2$ process. Any annihilation processes such as $\bar{\chi} \chi \to \bar{f} f $ are taken to be subdominant and assumed to decouple early. Additionally, there is a mediation process that maintains kinetic equilibrium between the dark and SM sectors, as in the SIMP mechanism. The key difference between ELDERs and SIMPs is the order in which processes decouple. In the SIMP paradigm, the DM-number-changing process stops while kinetic equilibrium is still maintained between the dark and SM baths. In the ELDER scenario, DM elastically decouples first, {\it i.e.}\/, the kinetic equilibrium between the two baths stops before the DM-number-changing process. In this case, when the DM elastically decouples ends up being the primary factor that determines the relic abundance. 

It is possible that the dark vector resonance we require to achieve the desired SIDM behavior is already realized in SIMP or ELDER scenarios. For concreteness, we consider one of the simplest SIMP realizations where the $3 \to 2$ process is realized by a Wess-Zumino-Witten term in a dark chiral Lagrangian where the dark pions compose DM~\cite{Hochberg:2014kqa}. Motivated by specific realizations of this SIMPlest miracle scenario~\cite{Hochberg:2018rjs}, we further restrict our consideration to an Sp(4) gauge group with $N_f=2$ fermions in the fundamental, so that the flavor symmetry is SU(4)/Sp(4) and there are 5, equal-mass pions comprising DM. 

For this gauge and flavor structure, there exist lattice results for the corresponding spectra and decay constants after confinement in the dark sector~\cite{Bennett:2019cxd}. In particular, there exists a single point at which the lightest pseudoscalar mass, {\it i.e.}\/, the dark pion, is exactly half the mass of the lightest vector resonance. At this point, the ratio of the pseudoscalar mass to its decay constant is~\footnote{Our convention for $f_K$ matches that of Ref.~\cite{Hochberg:2014kqa}, which is a factor of 2 larger than the convention used in Ref.~\cite{Bennett:2019cxd}.}
\al{
m_\pi / f_\pi = 1.9.
}
At first glance for this ratio, we find that the SIMP mechanism does not quite work as the necessary $m_K$~\cite{Hochberg:2014kqa} causes the DM self interaction to be too large and excluded by the Bullet Cluster bound~\cite{Clowe:2003tk,Markevitch:2003at,Randall:2007ph}. However, to see whether this parameter set simultaneously explains both the abundance and the self-interaction cross section requires detailed modeling of pion scattering including the vector meson exchanges, which is beyond the scope of this Letter and will be discussed elsewhere~\cite{New:2020}. Given a variety of QCD-like gauge theories, we believe a significant fraction of them leads to the correct phenomenology.

\section{Conclusion}

In this letter, we have presented various realizations of resonant self-interacting dark matter, using pseudoscalar and vector meson states arising from a dark QCD.
In particular, we have proposed a model in which the pseudoscalar mesons are composed of light quarks, analogous to the $\phi$ -$K$-$K$ system in the Standard Model, and considered the freeze-out scenario through a kinetically mixed dark photon.
We have also detailed a model of asymmetric dark matter in which the dark pseudoscalar-vector meson system is similar to the $\Upsilon(4S)$-$B$-$B$ system and the requisite dark photon is imminently discoverable at current and future experiments. Finally, we used the latest lattice results to consider the built-in resonant self-interaction which could be present in models of SIMP or ELDER dark matter. 
We expect the models we have presented in this letter to motivate future small-scale studies \cite{NewAstro:2020}, as well as experimental searches, to test the parameter space of interest.

\begin{acknowledgements}
We thank Asher Berlin, Xiaoyong Chu, Camilo Garcia-Cely, Manoj Kaplinghat, Gordan Krnjaic, Yuhsin Tsai, and Sean Tulin for useful discussions.

The work of RM was supported by NSF grant PHY-1915314 and the U.S. DOE Contract DE-AC02-05CH11231. The work of HM was supported by the NSF grant PHY-1915314, by the U.S. DOE Contract DE-AC02-05CH11231, by the JSPS Grant-in-Aid for Scientific Research JP17K05409, MEXT Grant-in-Aid for Scientific Research on Innovative Areas JP15H05887, JP15K21733, by WPI, MEXT, Japan, and Hamamatsu Photonics, K.K.

Part of this document was prepared by Y.-D.T. using the resources of the Fermi National Accelerator Laboratory (Fermilab), a U.S. Department of Energy, Office of Science, HEP User Facility. Fermilab is managed by Fermi Research Alliance, LLC (FRA), acting under Contract No. DE-AC02-07CH11359. Part of this work was performed by Y.-D.T. at the Aspen Center for Physics, which is supported by the National Science Foundation grant PHY-1607611.

\end{acknowledgements}

\bibliography{draft1}{}
\bibliographystyle{utcaps_mod}
\appendix

\section{SM-dark QCD meson summary}
Here, we first discuss the resonance structure in the theory of QCD and meson bound states. This is directly shown in Fig.~\ref{fig:resonance} and the details are described in the caption. In Table \ref{table:1}, we consider the vector meson to pseudoscalar meson couplings for four meson systems as examples. We consider the interaction 
\begin{align}
\mathcal{L_{\rm int}}=
g_V V^\mu (PS) \partial_\mu (\overline{PS}).
\end{align}
$g_V$ is the coupling, $V$ is the vector meson, and PS is the pseudoscalar meson. We also discuss the $\gamma$ parameter considered in \cite{Chu:2018fzy} for the purpose of fitting to data to determine the velocity dependence of the DM self-interaction, and determine the range of DM mass reading of the upper-right panel of Fig. 2 of \cite{Chu:2018fzy}.

\begin{table}[h!]
\centering
\begin{tabular}{|c | c | c | c || c| } 
 \hline
 systems & $g_V$ & $\gamma = \frac{g_V^2}{(384\pi)}$
& $\Delta\equiv 1-\frac{2 m_{\rm PS}}{m_{\rm V}}$& $m_{\rm DM}$ [GeV]  \\ [1ex] 
 \hline
$\phi$-$K$-$K$ & 4.5 & ~0.02~ &
0.02 & *0.9 - 1.5   \\ 
\hline
$\Upsilon (4S)$-$B$-$B$
& ~25~ & 0.50& $2 \times 10^{-3}$ & *2.8 - 4.7 \\
 \hline
\end{tabular}
\caption{
In this table, we show two vector-to-pseudoscalar meson systems that inspired our models. $g_V$ is the coupling between the vector mesons to pseudoscalar mesons defined in the text, and $\gamma$ is considered in \cite{Chu:2018fzy} and fitted to the small-scale structure data to determine the best range of DM masses that yield the desired velocity dependence of the self-interactions. Note that for the $\phi$-$K$-$K$ system here we use $K^{\pm}$ instead of $K^{0}$, to better match the discussions in Sec. \ref{sec:light-quark}. A “*” indicates that, to get these $m_{\rm DM}$ ranges, we consider $\Delta \sim 10^{-7.8}$ instead of the SM values, to have the desired resonant self-interaction.
}
\label{table:1}
\end{table}

\section{More details on the light quark model}
\label{app:A}

\begin{figure}[h!]
    \centering
    \includegraphics[
    width=7cm]{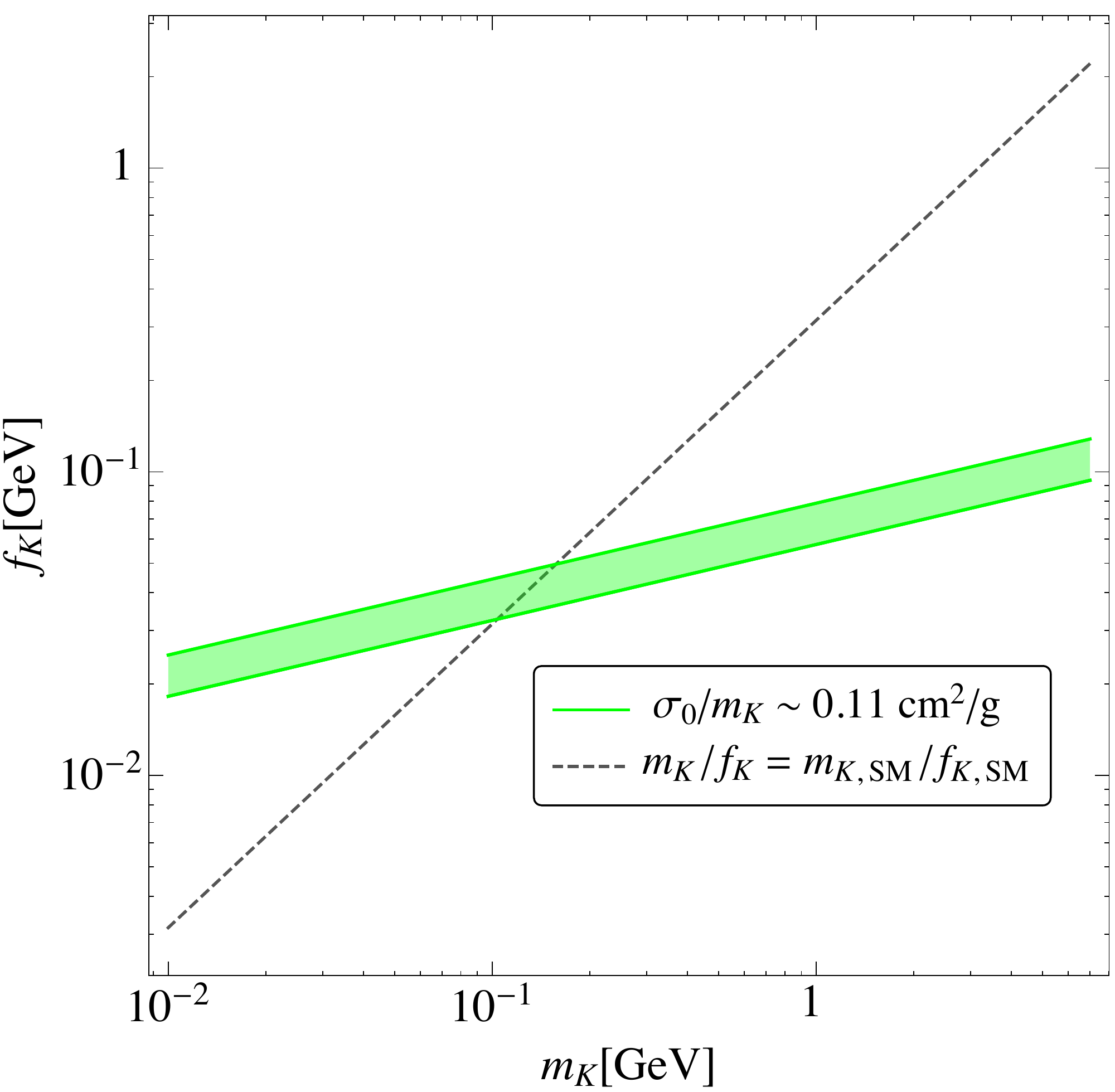}
    \caption{We show in green the dark-kaon masses and decay constants which yield the correct self-interaction cross-section in the low-velocity limit. The dotted curve follows the SM ratio, $m_{K}/f_{K}$ = $m_{K, \rm SM}/f_{K, \rm SM}$.
    }\label{fig:mvf}
\end{figure}

In Fig.~\ref{fig:mvf}, the green band shows $m_{K}$ versus $f_K$ values which yield the correct self-interaction cross section in the low-velocity limit for the light-quark model in Sec. \ref{sec:light-quark}. The dotted curve corresponds to $m_{\rm K}$ versus $f_K$ values which equal the SM ratio with $m_{K}/f_{K}$ = $m_{K,\rm SM}/f_{K, \rm SM}$.

{\bf Changing $N_c$ for the light-quark model - }Following the discussion in section \ref{sec:light-quark}, we discuss the implications of changing the number of colors $N_c$ for the $SU(N_c)$ gauge group. Note that the dependence of hadronic parameters on $N_c$ can be discussed reliably only in the large $N_c$ limit, and applying it to $N_c \sim 3$ is only qualitative and suggestive at best.

For the consideration of changing $N_c$, we fix the parameter $\Delta$, and thus the $m_V/m_{\rm PS}$. Again, $V$ is the vector resonance, and PS denotes the pseudoscalar DM, so that the resonant self-interaction is unaffected.
We also fix several ratios of the parameters while conducting the $N_c$ scaling.
First, we fix $m_{K}/\Lambda$, and $\Lambda$ is the scale of the dark QCD, and we also fix $m_q/\Lambda$.
We set all the mass ratios to the SM ratios when $N_c = 3$, and then we change $N_c$ to see the behavior of the change,
\begin{align}
\frac{m_{K}}{f_K} = \left(\frac{m_{K, \rm SM}}{f_{K, \rm SM}}\right) \sqrt{\frac{3}{N_c}}.
\end{align}

Now, for region I, we again apply the condition Eq. \ref{eq:sigma_0}, but now we have to modify $\sigma_0/m_{\rm DM}$ as
\begin{align}
\begin{split}
\frac{\sigma_0}{m_{\rm DM}} &= \frac{1}{32 \pi} \left(\frac{m_K^4}{f_K^4}\right)_{\rm }\frac{1}{m_K^3} \\
&= \frac{1}{32 \pi} \left(\frac{m_{K, \rm SM}}{f_{K, \rm SM}}\right)^4 \left(\frac{3}{N_c}\right)^2 \frac{1}{m_{K}^3} .\\
\end{split}
\end{align}
Since we are fixing the value of $\sigma_0/m_{\rm DM}$ and $m_{K, \rm SM}/f_{K, \rm SM}$ is taken from data, $\left(\frac{3}{N_c}\right)^2 \frac{1}{m_{K}^3}$ is a constant. Thus, we arrived that $m_{K} = m_{{K}, N_c = 3}\left(\frac{3}{N_c}\right)^{2/3}$ for the region I dark kaon mass range to scale with $N_c.$

For region II, the argument is much simpler. Since $g_V \propto \sqrt{\frac{1}{N_c}}$, we find $\gamma\propto \frac{1}{N_c}$. Based on \cite{Chu:2018fzy}, we have
$m_K\propto \gamma^{1/3}$, so $m_{K} = m_{{K}, N_c = 3}\left(\frac{3}{N_c}\right)^{1/3}$ when one scales the region II parameter according to different $N_c$.

One can see that, based on this simple naive argument, regions I and II move closer to each other for small $N_c$. However, as we know the large $N_c$ expression breaks down for small $N_c$, so the above scaling is just a suggestive consideration. Nonetheless, it may indicate that the two regions could coincide in small $N_c$, and motivate the future lattice study.

Note that other gauge groups could work better for this purpose.  For instance, an $Sp(2N_c)$ gauge theory with $N_f=2$ (four Weyl fermions in the fundamental representation) would lead to chiral Lagrangian with the coset space $SU(4)/Sp(4) = SO(6)/SO(5) = S^5$.  Vector mesons are in the adjoint representation of $SO(5)$.  Upon gauging a $U(1)$ subgroup, the symmetry breaks to $U(2)$ and the Nambu--Goldstone bosons split as $(K^+, K^-, K_1^0, K_2^0, K_3^0)$.  When the neutrals are heavier, the low-lying spectrum is the same as the $SU(N_c)$ gauge theories we discussed, while one of the vector mesons can appear in the $K^+ K^-$ channel.  Similarly, an $SO(N_c)$ gauge theory with $N_f=2$ (two Weyl fermions in the vector representation) has the coset space $SU(2)/SO(2)=S^2$ which comprises with only $K^\pm$.  We expect a single vector meson in the $K^+ K^-$ channel in this case as well.  Quantitative discussions on such possibilities are beyond the scope of this paper and will be discussed in \cite{New:2020}.

\end{document}